# Wavelet Analysis of Cycle-to-Cycle Pressure Variations in an Internal Combustion Engine

Asok K. Sen

Department of Mathematical Sciences

Indiana University

402 N. Blackford Street

Indianapolis, IN 46202, USA

Email: asen@iupui.edu

Grzegorz Litak

Department of Mechanics

Technical University of Lublin

Nadbystrzycka 36, PL-20-618, Lublin, Poland

Email: g.litak@pollub.pl

Rodolfo Taccani

Department of Mechanical Engineering

University of Trieste

Via A. Valerio 10, I-34127, Trieste, Italy

Email: taccani@univ.trieste.it

and

Robert Radu

Department of Mechanical Engineering

University of Trieste

Via A. Valerio 10, I-34127, Trieste, Italy

Email: rradu@univ.trieste.it

# ABSTRACT

Using a continuous wavelet transform we have analyzed the cycle-to-cycle variations of pressure in an internal combustion engine. The time series of maximum pressure variations are examined for different loading and their wavelet power spectrum is calculated for each load. From the wavelet power spectrum we detected the presence of long, intermediate and short-term periodicities in the pressure signal. It is found that depending on the load, the long and intermediate-term periodicities may span several cycles, whereas the short-period oscillations tend to appear intermittently. Knowledge of these periodicities may be useful to develop effective control strategies for efficient combustion.

# 1. Introduction

For many years there has been considerable interest in studying the cycle-to-cycle changes of the process variables such as pressure in an internal combustion engine [1-10]. These studies are important to gain a better understanding of the mechanisms by which the various factors affect the overall combustion process. The factors include the fuel-air ratio in the combustible mixture, amount of recycled gases supplied to the cylinder, and engine aerodynamics. The information on cylinder pressure variations can be used to control engine combustion [2]. Recent efforts have focused on considering the cycle-to-cycle variability in the realm of nonlinear dynamics and applying chaos-theoretic methods to unravel the nonlinear aspects of these variations [3-7]. It has also been shown that the variability may be due to a stochastic component [8], and attempts have been made to estimate the noise level [9, 10]. The purpose of this paper is to apply wavelet-based techniques to analyze the cycle-to-cycle variations of maximum pressure in an internal combustion engine. In particular, we use a continuous wavelet transform on the time series of maximum pressure variations under different loading and calculate the wavelet power spectrum to detect the presence of long, intermediate and short-term periodicities and intermittency in the pressure signal. We find that depending on the load, the long and intermediate-term periodicities may span several cycles, whereas the short-term periodicities tend to appear intermittently. It should be pointed out that although it is possible to use a Fourier transform to detect many of these periodicities, a Fourier transform cannot determine their temporal variations because it is a purely frequency domain approach and thus it loses all the temporal information in a signal. A simple way to capture the temporal aspects is to analyze a short segment of the signal at a time using a window of fixed size and then sliding the window along in time. This procedure is known as the Short-Time Fourier Transform (STFT) or Windowed Fourier Transform (WFT). Because STFT/WFT uses a fixed-size window, it suffers from the following drawback: at high periodicities it has a poor frequency resolution while at low periodicities the time resolution is poor. As a consequence, an STFT/WFT may give an inconsistent estimation of the periodicities and their temporal variations for a signal with multiple frequencies. Wavelet analysis provides an efficient approach by which both the time and frequency resolutions can be adjusted in an adaptive fashion. It offers high frequency resolution and low temporal resolution at high periodicities, and high temporal resolution and low frequency resolution at low periodicities. Knowledge of these periodicities may be useful to develop effective control strategies for efficient combustion. Wavelets have been used for signal analysis in a wide variety of applications [11-13]. Some researchers have used wavelet transforms to detect hidden periodicities and intermittency in experimentally measured and observed time series (see, for example, [14-17]).

## 2. Data Acquisition

All the data analyzed in this paper were collected from a four-stroke, single-cylinder Aprilia/Rotax spark ignition engine. The experiments were performed in the Engine Laboratory at the University of Trieste, Italy. Details of the experimental design are given in an earlier paper [10].

For the purpose of acquiring data, the torque on the engine was varied and the internal pressure in the engine cylinder was measured for 1000 cycles with 6000 points per cycle for each torque loading. From these measurements the maximum value of pressure ($p_{\max}$) in each cycle is found for a given load. The time series of cycle-to-cycle variations of $p_{max}$ for the first 960 cycles are plotted in Figure 1, for torques F = 0, 10, 20, 28, 40 and 43 Nm. In the following development we perform a wavelet analysis of these time series to determine the various periodicities of maximum pressure fluctuations.

## 3. Wavelet Analysis

A wavelet is a small wave with a compact support. In order to be classified as a wavelet, a function $\psi(t)$ should have zero mean and finite energy:

$$\int_{-\infty}^{\infty} \psi(t)\, dt = 0, \qquad \int_{-\infty}^{\infty} |\psi(t)|^2\, dt < \infty. \tag{1}$$

The continuous wavelet transform (CWT) of a function $x(t)$ with respect to a wavelet $\psi(t)$ is defined as a convolution of the function with a scaled and translated version of $\psi(t)$. The wavelet $\psi(t)$ is referred to as an analyzing wavelet or a mother wavelet. The convolution is expressed by the integral [11]:

$$W(s,\tau) = \int_{-\infty}^{\infty} x(t)\psi^*_{s,\tau}(t)\, dt, \tag{2}$$

where

$$\psi_{s,\tau}(t) = \frac{1}{\sqrt{s}} \psi\left(\frac{t-\tau}{s}\right) \tag{3}$$

is a scaled and translated version of the mother wavelet $\psi(t)$, and an asterisk on $\psi$ denotes its complex conjugate. The symbols $s$ and $\tau$ are called a scale parameter and a translation parameter, respectively. The scale parameter controls the dilation ($s > 1$) and contraction ($s < 1$) of the wavelet, whereas the translation parameter indicates the location of the wavelet in time. The factor $1/\sqrt{s}$ is introduced in Eqn (3) so that $\psi_{s,\tau}(t)$ has unit energy at every scale. The wavelet power spectrum (WPS) of the signal representing the energy at scale *s* is defined as the squared modulus of the CWT:

$$P(s,\tau) = |W(s,\tau)|^2. \tag{4}$$

The integral formulation for a CWT as given in Eqn (2) applies to a signal *x(t)* that is a continuous function of *t*. In order to use it for a discrete signal such as the pressure signal, this integral representation must be discretized in an appropriate fashion. Consider a discrete time series $\{x_n\}$ with $n = 1, 2, 3, …, N$. For such a time series, Eqn (2) may be discretized as [18]:

$$W_n(s) = \sum_{n'=1}^{N} \left(\frac{\delta\tau}{s}\right)^{1/2} x_{n'} \psi^*\left[\frac{(n'-n)\delta\tau}{s}\right]. \tag{5}$$

Here *n* is the time index, and $\delta\tau$ is the sampling interval. The factor $(\delta\tau/s)^{1/2}$ preserves the unit energy property referred to earlier. In order to calculate the CWT using Eqn (5), the convolution procedure given by this equation should be performed *N* times for each scale. However, as shown in [18], it is possible to carry out all *N* convolutions simultaneously in Fourier space by means of a discrete Fourier transform (DFT).

For a time series of finite length, computation of CWT using DFT requires that the time series is cyclic. To satisfy this requirement, the time series is often padded with zeros. Zero padding leads to errors at the ends of the wavelet power spectrum. The region in which these edge effects become important is called the cone of influence (COI). As a consequence, the results outside the COI may be unreliable and should be interpreted with caution. In the figures presented here, the COI is indicated by a thin solid curve.

Using Eqn (5), the WPS for each scale can be evaluated as: $|W_n(s)|^2$. The WPS provides a good description of the fluctuation of the variance at different scales or frequencies. It is customary to normalize the WPS by $\sigma^{-2}$, $\sigma$ being the variance. The normalized WPS thus gives a measure of the power relative to white noise. The WPS which depends on both scale and time is represented by a surface. By taking contours of this surface and plotting them on a plane, a time-scale representation of the wavelet power spectrum may be derived. The time-scale representation is also referred to as a scalogram. From a scalogram the various periodicities and intermittency can be identified by visual inspection. In our analysis we used a complex Morlet wavelet as the mother wavelet. A Morlet wavelet consists of a plane wave modulated by a Gaussian function and is described by:

$$\psi(\eta) = \pi^{-1/4} e^{i\omega_0 \eta} e^{-\eta^2/2}, \qquad (6)$$

where $\omega_0$ is the center frequency, also referred to as the order of the wavelet. Morlet wavelets have been used successfully in a wide variety of applications for feature extraction in time series data. In our analysis we have used a Morlet wavelet with $\omega_0 = 6$. This value of $\omega_0$ provides a good balance between time and frequency localizations. For $\omega_0 = 6$, the Fourier period is equal to 1.03 times the scale $s$. Note that in Eqn (6) the coefficient $\pi^{-1/4}$ is used as a normalization factor to ensure that the Morlet wavelet has unit energy.

In addition to the WPS which is local in nature, additional information about the behavior of a time series can be obtained by calculating the scale-averaged wavelet power (SAWP). The scale-averaged wavelet power represents the average variance in a certain range of scales (or a band). The SAWP over a specific band represents the average variance in the band; in other words, it describes the fluctuations in power in that band. In particular, by plotting the variations of SAWP, the presence of intermittency, if any, can be clearly identified. The SAWP is defined as the weighted sum of the wavelet power spectrum over scales corresponding to $j_1$ and $j_2$:

$$\overline{W_n}^2 = = \frac{\delta_j \delta\tau}{C_\delta} \sum_{j=j_1}^{j=j_2} \frac{|W_n(s_j)|^2}{s_j}. \tag{7}$$

Here $\delta\tau$ is the sampling interval as indicated earlier, and the factor $\delta j$ determines the scale resolution. The constant $C_\delta$ is a reconstruction factor whose value depends on the choice of the specific mother wavelet; for a Morlet wavelet with $\omega_0 = 6$, $C_\delta = 0.776$ (see [18] for details).

## 4. Results

Figure 2 depicts a time-scale representation of the wavelet power spectrum (WPS) of the pressure signal for no loading. Several periodicities can be readily observed in this figure. For example, there are periodic bands bordering the cone of influence (COI). Among these bands, the 165-265 cycle band has the strongest intensity. Another strong periodicity consists of the 52-66 cycle band and it persists from approximately 370 to 460 cycles. A less strong and narrower periodic band is the 9-14 cycle band and it spans 420 to 440 cycles. In addition, there are several weaker short-term periodicities covering the 2-8 cycle band that appear intermittently.

A time-scale representation of the wavelet power spectrum of the pressure signal for a load of F = 10 Nm is shown in Figure 3. In comparison to the scenario in Figure 2 (with no load), the strong periodic bands in this figure bordering the COI extends to smaller periodicities, indicating higher frequencies produced by the application of loading. The strongest periodicity is now confined to the 185-265 cycle band. Two weaker periodic bands with (a) 57-63 cycle period spanning approximately 750 to 850 cycles, and (b) 33-37 cycle band occuring between 545 to 585 cycles are also seen in this figure. Furthermore, a few weaker intermittent short-term periodicities are present in Figure 3.

The results for F = 20 Nm are illustrated in Figure 4. As seen in this figure, an increase in torque from 10 Nm to 20 Nm has reduced the periodic band bordering the COI and the reduced band tends to appear more uniform. A strong distinct 75-120 cycle band spanning approximately from 120 to 160 cycles along the border to 375 cycles is also visible in this figure. In addition, we observe a few weaker periodicities and several short-term periodicities that are intermittent.

A scalogram of the maximum pressure variations for a torque of 28 Nm is shown in Figure 5. This figure reveals that as the torque is increased from 20 Nm to 28 Nm, the periodicities bordering the COI have almost completely disappeared except for a very small band around the 64-cycle. We now see a 28-40 cycle band over the span of 815 to 895 cycles. In addition, there are a few short-term periodicities in the 8-16 cycle range and several even shorter-term periodicities that are intermittent.

Figure 6 presents the results for F = 40 Nm. The two most prominent periodicities in this figure are: (a) 95-175 cycle band between 250 to 640 cycles, and (b) 60-80 cycle band from 135 to 305 cycles. Other significant periodicities include: (c) 18-30 cycle periodicity over 405 to 470 cycles, and (d) 7-18 cycle periodicity ranging from 115 to 160 cycles. A few short-term intermittent periodicities are also present in this figure.

Finally, the results for the load of 43 Nm are presented in Figure 7. The following periodicities are apparent in this figure. Bordering the COI, there are: (a) 120-210 cycle band, and (b) 37-70 cycle band. In addition, we see (c) 45-70 cycle periodic band from 570 to 690 cycles, (d) 13-25 cycle band between 815 and 880 cycles, and (c) 15-24 cycle band spanning 50 to 90 cycles. There are also several short-term periodicities appearing in an intermittent fashion.

It was mentioned in Section 3 that in addition to a wavelet power spectrum (WPS), additional information about the behavior of a time series can be obtained from the computation of scale-averaged wavelet power (SAWP), particularly in regard to intermittency [16, 17]. To illustrate the use of SAWP for detecting intermittency, we consider the time series of maximum pressure variations with a torque of 20 Nm. For this time series, Figure 8(b) shows the variation of scale-averaged power in the 2-8 cycle band revealing the intermittent nature of these oscillations. To put things in perspective, we have also plotted the scale-averaged wavelet power for the 75-120 cycle periodic band. This is displayed in Figure 8(c). It is clear from this figure that in this periodic band the average variance is very low between approximately 400 and 700 cycles.

## 5. Discussion and Conclusion

Using a continuous wavelet transform we have analyzed the cycle-to-cycle variations of maximum pressure in an internal combustion engine. Several long, intermediate and short-term periodicities have been identified from the wavelet power spectrum of the pressure signal.

From a comparison of Figures 2 through 7, the following conclusions can be drawn. In the absence of loading (F = 0), and for small loading (F = 10), there is a prominence of longer-term periodicities, and they persist over many cycles. At higher loads, these periodicities tend to become weaker or disappear altogether, and several intermediate and short-term periodicities appear. Many of the short-term periodicities are intermittent in nature.

One should note that the combustion process in the engine cylinder influences the maximum pressure $p_{max}$ (Fig. 1) through the combined effects of mechanical compression of the cylinder gas volume caused by a cyclic motion of the piston and chemical combustion [10]. In the case of small loading, chemical combustion is minimized by adding relatively smaller portions of fuel. In this region of engine work (small loading: F = 0 or 10 Nm) the combustion process itself may be incomplete or even absent (misfire) due to inhomogeneous distribution of the fuel in the engine chamber (especially in the neighborhood of the spark plug electrodes). Thus the longer-term periodicities visible in Figs. 2 and 3 mimic the cyclic pressure compression maxima perturbed by non-periodic combustion. On the other hand, the cases with higher loading (e.g., F = 28, 40, 43 Nm) are dominated by a combustion process with nontrivial residual gas mixing [1, 3] with new portions of fresh air and fuel at each combustion cycle. The fingerprint of such behavior is the appearance of short-term periodicities. The case F = 20 Nm is an intermediate situation where processes of both compression and chemical combustion may be equally important to create the maximum pressure $p_{max}$ (Fig. 1), and an intermittent behavior [7] appears on the basis of the two competing phenomena as seen in Figs. 8a-b. Special attention should be paid to the case F = 28 Nm which does not match the others (see, for example, Fig. 1 and Fig. 5) because of the absence of long- and intermediate-term periodicities. This could be caused by a development of chaotic combustion [3, 5, 6]. To check such a possibility one should perform further investigations by looking for an embedding dimension and calculating the largest Lyapunov exponent [19, 20].

.

# Figure Legends

Figure 1.  Time series of cycle-to-cycle variations of maximum pressure for loads F = 0, 10, 20, 28, 40 and 43 Nm.

Figure 2.  Time-scale representation of the wavelet power spectrum of the time series of the pressure signal with no load as shown in Figure 1, obtained by a continuous wavelet transform. A Morlet mother wavelet with $\omega_0 = 6$ is used. The thick contour lines represent the 5% significance level and the lighter curve denotes the cone of influence (COI).

Figure 3.  Time-scale representation of the wavelet power spectrum of the time series of the pressure signal for a load F =10 Nm as shown in Figure 1.

Figure 4.  (a) Time-scale representation of the wavelet power spectrum of the time series of the pressure signal for F = 20 Nm as shown in Figure 1.

Figure 5.  Time-scale representation of the wavelet power spectrum of the time series of the pressure signal for F = 28 Nm as shown in Figure 1.

Figure 6.  Time-scale representation of the wavelet power spectrum of the time series of the pressure signal for F = 40 Nm as shown in Figure 1.

Figure 7.  Time-scale representation of the wavelet power spectrum of the time series of the pressure signal for F = 40 Nm as shown in Figure 1.

Figure 8.  Scale-averaged wavelet power of the time series of the pressure signal for F = 20 Nm as shown in Figure 1. (a) averaged over the 2-8 cycle band, and (b) averaged over the 70-120 cycle band.

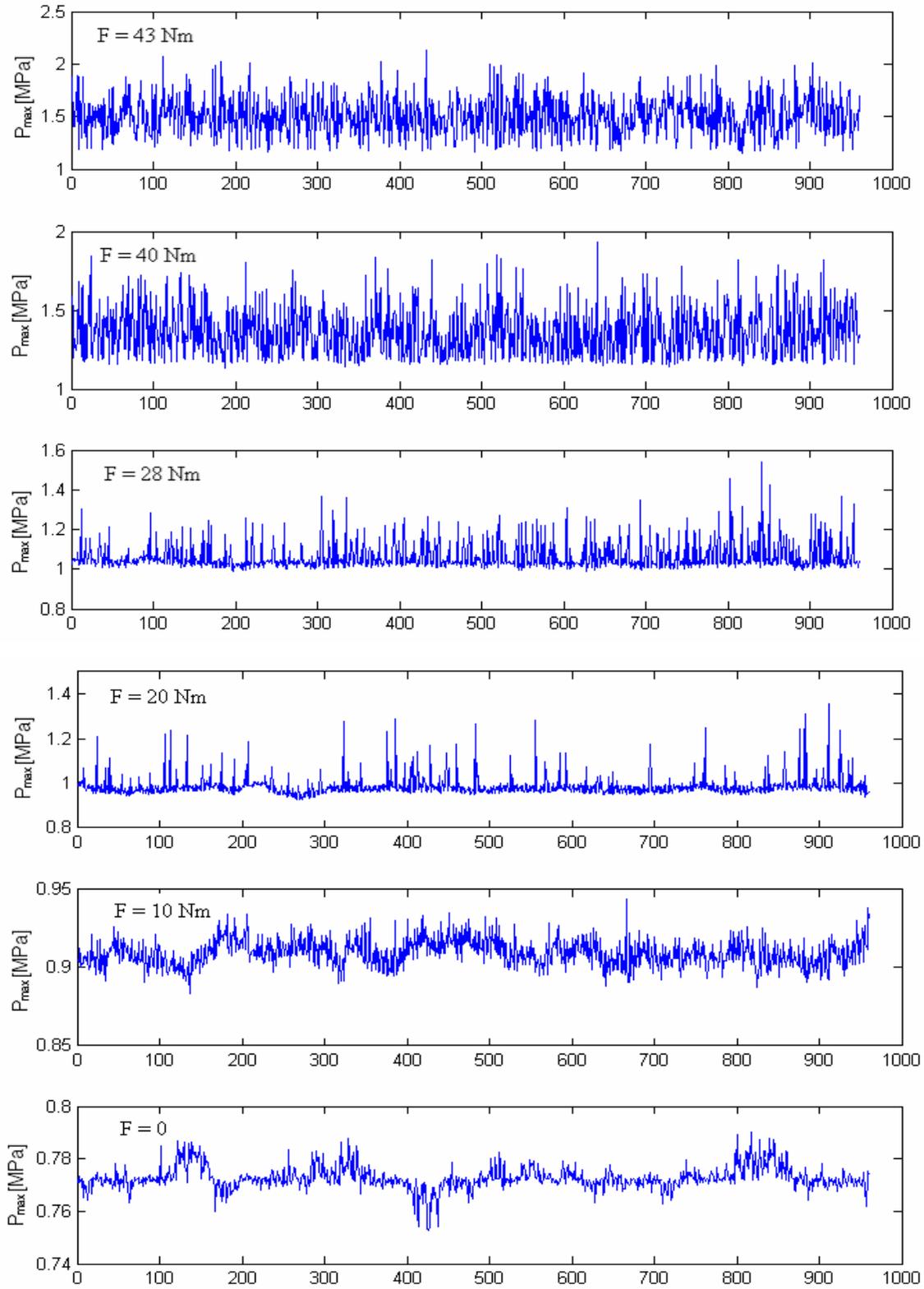

Figure 1

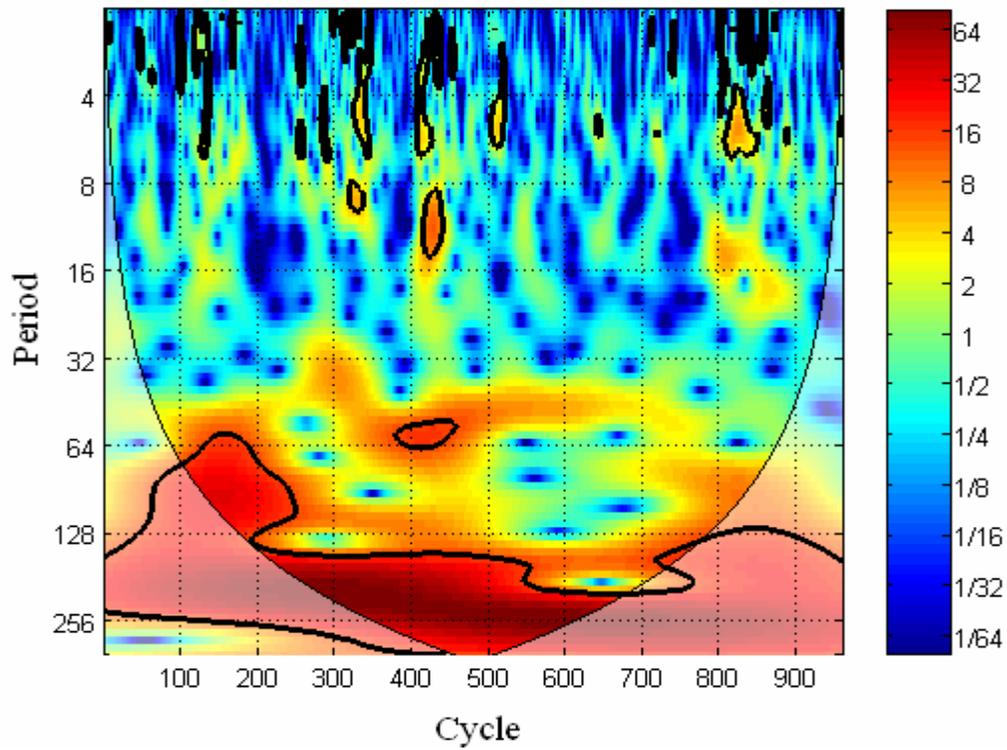

Figure 2

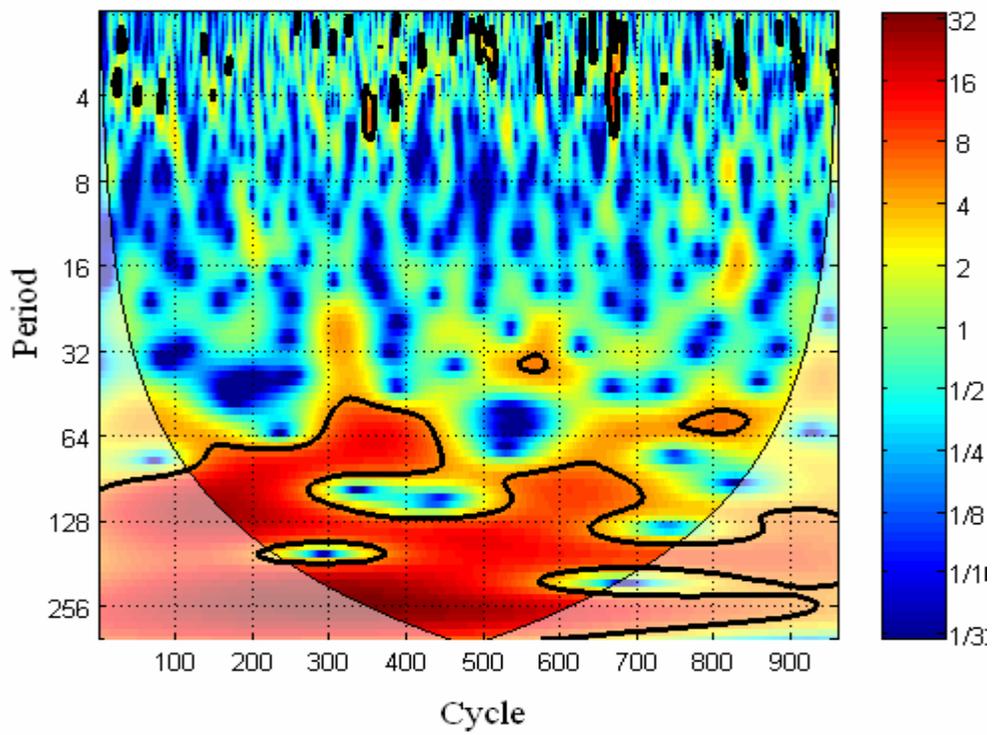

Figure 3

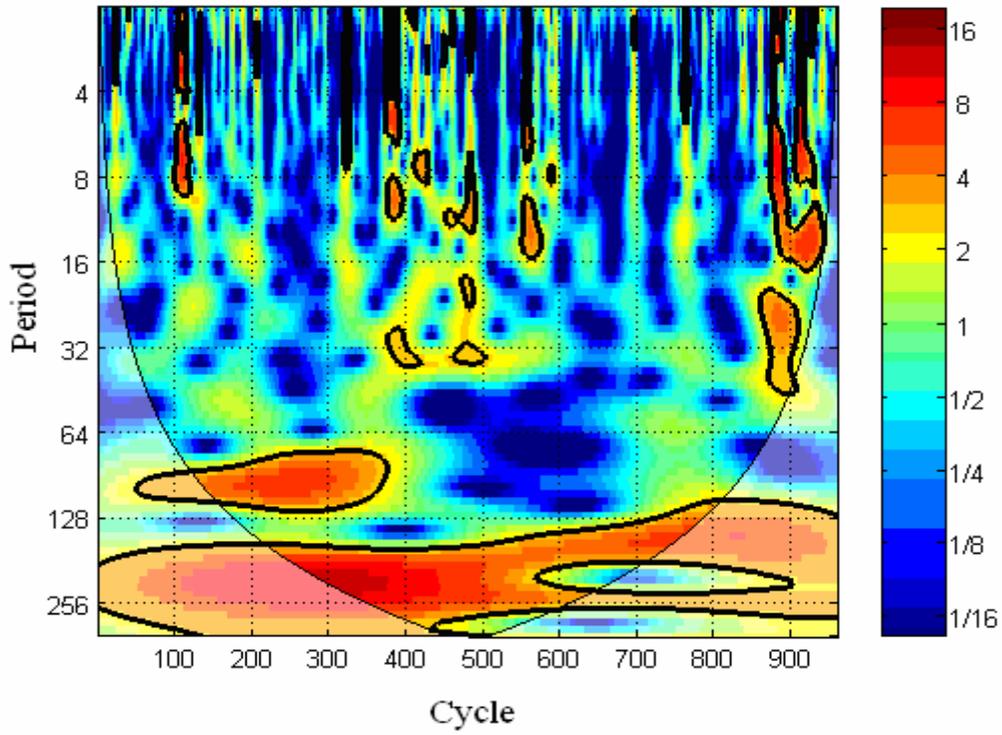

Figure 4

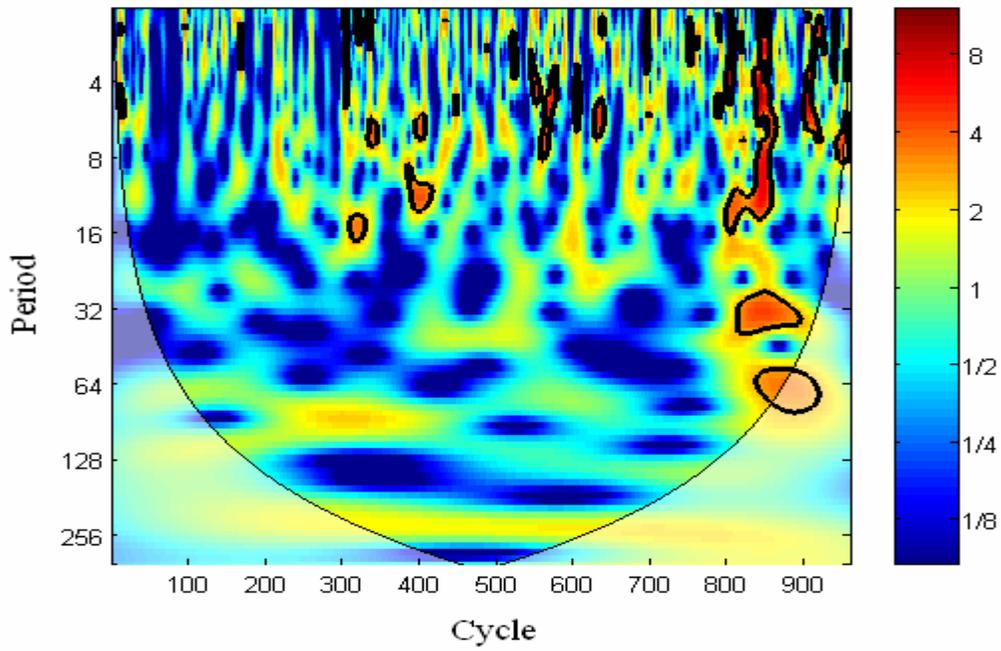

Figure 5

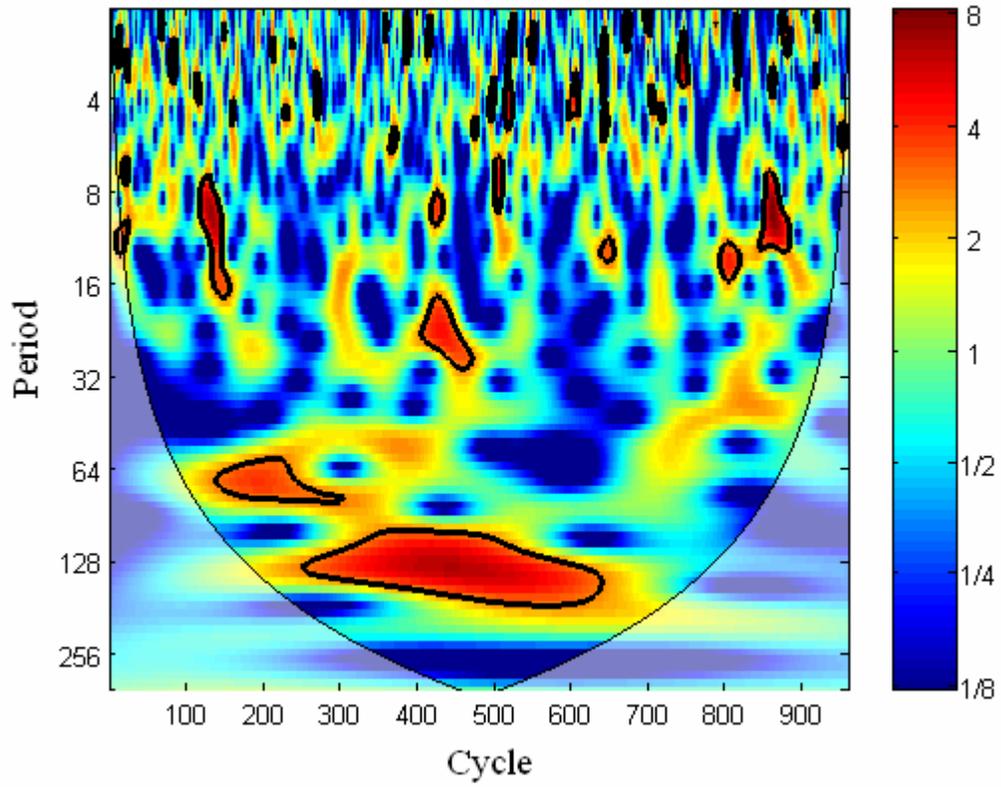

Figure 6

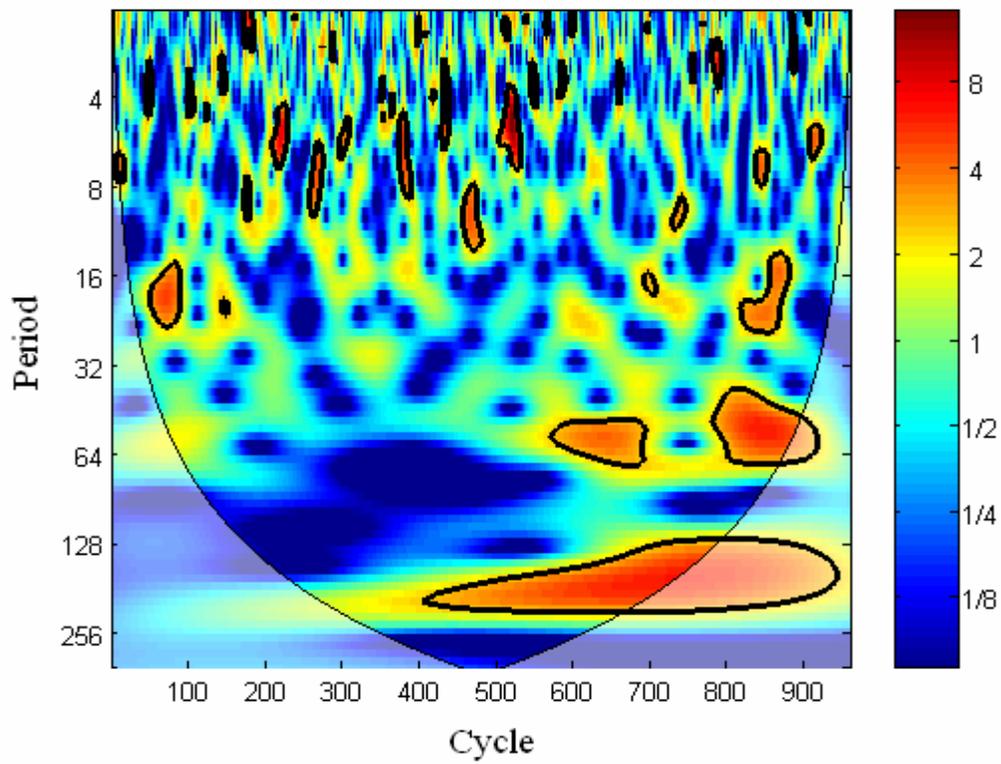

Figure 7

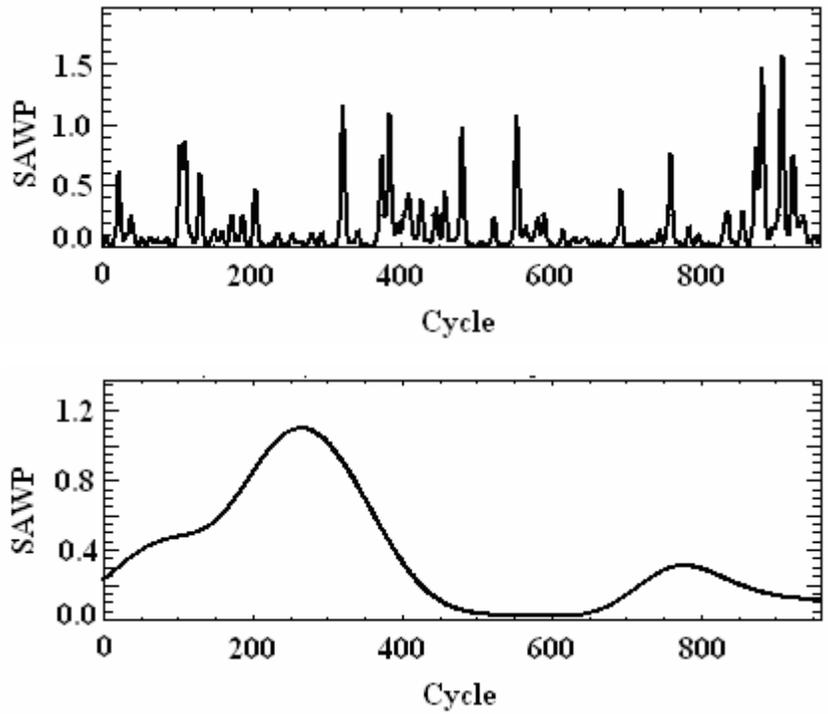

Figure 8